\documentclass[preprint,5p,10pt]{elsarticle}

\usepackage{graphicx}
\usepackage{dcolumn}
\usepackage{bm}
\usepackage{amssymb}
\usepackage{amsmath}
\usepackage{color}

\journal{105 Group Science}
\begin{document}

\begin{frontmatter}
  
\title{When climate variables improve the dengue forecasting: a machine learning
approach}

\author{Sidney T. da Silva$^{1,*}$, Enrique C. Gabrick$^{2,3,4,*}$, 
Paulo R. Protachevicz$^{5}$, Kelly C. Iarosz$^{2,6}$, 
 Iber\^e L. Caldas$^{5}$, Antonio M. Batista $^{2,7}$, 
Jürgen Kurths$^{3,4}$}
\address{$^1$Federal University of Paran\'a, 81531-980, Curitiba, PR, Brazil.\\}
\address{$^2$Graduate Program in Science, State University of Ponta Grossa, 84030-900, Ponta Grossa, PR, Brazil.\\}
\address{$^3$Department of Physics, Humboldt University Berlin, Newtonstra{\ss}e 15, 12489 Berlin, Germany.\\}
\address{$^4$Potsdam Institute for Climate Impact Research, Telegrafenberg A31, 14473 Potsdam, Germany.\\}
\address{$^5$Institute of Physics, University of S\~ao Paulo, 05508-090, S\~ao Paulo, SP, Brazil.\\}
\address{$^6$University Center UNIFATEB, 84266‐010, PR, Tel\^emaco Borba, Brazil.\\}
\address{$^7$Department of Mathematics and Statistics, State University of Ponta Grossa, 84030-900, Ponta Grossa, PR, Brazil.\\}

\cortext[cor]{sidneyt.silva82@gmail.com (STS); ecgabrick@gmail.com (ECG)}

\begin{abstract}{Dengue is a viral vector borne infectious disease that affects many
countries worldwide, infecting around 390 million people per year. The main 
outbreaks occur in subtropical and tropical countries. 
We, therefore, study here the influence of climate on dengue. 
In particular, we consider
dengue and meteorological data from Natal (2016-2019), Brazil, Iquitos
(2001-2012), Peru, and Barranquilla (2011-2016), Colombia. 
For the analysis and simulations, we apply Machine Learning (ML) techniques, 
especially the Random Forest (RF) algorithm. 
We utilize dengue disease cases and climate data delayed by up to one week to forecast the cases of
dengue. In addition, regarding as feature in the ML technique, we analyze three
possibilities: only dengue cases (D); climate and dengue cases (CD);  
humidity and dengue cases (HD). Depending on the city, our results show that the
climate data can improve or not the forecast. For instance, for Natal, the
case D induces a better forecast. For Iquitos, it is better to use all the
climate variables. Nonetheless, for Barranquilla, the forecast is better, when we
include cases and humidity data. Another important result is that each city has
an optimal region based on the training length. For Natal, when we use more than
64\% and less than 80\% of the time series for training, we obtain results with
correlation coefficients ($r$) among 0.917 and 0.949 and mean absolute errors
(MAE) among 57.783 and 71.768 for the D case in forecasting. 
The optimal range
for Iquitos is obtained when 79\% up to 88\% of the time series is considered for training.
For this case, the best case is CD, having a minimum $r$ equal to 0.850 
and maximum 0.887, while values of MAE oscillate among 2.780 and 4.156.
For Barranquilla, the optimal range occurs between 72\% until 82\% of 
length training. In this case, the better approach is HD, where the measures 
exhibit a minimum $r$ equal to 0.942 and maximum 0.953, while the minimum and
maximum MAE vary among 6.085 and 6.669. We show that the forecast of dengue 
cases is a challenging problem and climate variables do not always help. However, 
when we include the mentioned climate variables, the most important one is the
humidity.}
\end{abstract}


\end{frontmatter}

%%%%%%%%%%%%%%%%%%%%%%%%%%%%%%%%%%
%%%%%%%%%%%%%%%%%%%%%%%%%%%%%%%%%%
\section{Introduction}
The number of dengue cases has increased over the past 2 decades. The world-wide 
reported cases in 2000 were 500,000 but have an incredible increase to
5.2 million in 2019 \cite{Lancet}. Most of these cases were reported by
countries in regions of Americas. For several countries belonging to these
region, the beginning of 2024 is marked by an exponential increase in the
dengue cases \cite{PAHO}. In regions of Americas, 4,565,911 cases were reported
in 2023, where 0.17\% were severe cases and 0.051\% implied in deaths. This
problem intensified in 2024, where 673.26 cases were reported already during 
the first 5 epidemiological weeks. From these amounts, 0.1\% were severe and 0.015\%
of the cases ended in deaths. This way, dengue is a challenging problem for 
public health and in particular for the dengue outbreak prediction models
\cite{Bhatt2013,WHO2023a}.

The dengue virus is transmitted to humans by bites of infected {\it Aedes}
mosquitoes, mainly by {\it aegypti} and {\it albopictus} \cite{Rosen1983,WHO}. 
There are four known dengue serotypes, which are designated by the acronym
followed by the index, as DENV-1, 2, 3 and 4 \cite{Messina2014}.Once infected by
one serotype, the individual acquires permanent immunity against it and
temporary cross-immunity to other ones \cite{Simmons2012}. Secondary infections
increase the probability of dengue hemorrhagic fever \cite{Halstead2003}.
Besides the consequences for human health, this disease has significant impacts
on the society and economy \cite{Shepard2016}. Although the dengue incidence is spatially 
distributed,  higher outbreaks occur in the tropical countries, mainly in 
South America \cite{ECDPC}. Some reasons for that  are due to the environmental conditions and climate
factors \cite{Morin2013}, such as temperature \cite{Watts1987} and precipitation
\cite{Reinhold2018}. The temperature acts directly in the life cycle and
reproductive of the mosquitoes \cite{Garin2000,Eisen2014}, while the rain
precipitation provides containers for eggs and habitat for the mosquitoes
\cite{Alto2001,Lega2017}. Moreover, the relationship between rainfall and
temperature are essential for regulating the water habits. Therefore, the
climate variables play an important role in the seasonality of dengue epidemics
\cite{Kraemer2019,Xavier2021}. 

The correlation between climate and dengue cases has been studied by many
approaches \cite{Xavier2021,Lowe2016,Oki2012,Paul2015,Johansson2009,Johansson2016,Yuan2020}, 
among them Machine learning (ML) techniques \cite{Carvajal2018}. 
Considering a multi-stage ML approach, Appice et al. investigated the evolution
of dengue cases in Mexico \cite{Appice2020}. To improve forecasting, they
analyzed the temperature influence. Guo et al., by means of  five ML
algorithms, developed a forecast model for dengue data from China
\cite{Guo2017}. Additionally, they took into account climate data, such as mean
temperature, relative humidity and rainfall. They showed that the support
vector regression algorithm is the best one to forecast the outbreaks in China. 
An ensemble neural network model was proposed to forecast dengue outbreaks based
on rainfall data from San Juan, Iquitos and Ahmedabad \cite{Panja2023}. The
authors used a framework available to forecast long-term cases around $52$
weeks. Deep learning techniques were employed by Zhao et al. to study and 
forecast the dengue cases in Singapore \cite{Zhao2023}. They also considered the
average temperature and rainfall time series. The best performance for $2$, $3$
and $4$ weeks in advanced forecasting was $84.61\%$. Other algorithms have been
employed to forecast dengue cases based on climate data \cite{Cabrera2022,Francisco2021,Salim2021,Rahman2021,Ochida2022,Ming2022}. 

Roster et al. used ML based on meteorological variables to forecast dengue cases
in Brazil \cite{Roster2022}. They considered Random Forest (RF), gradient
boosting regression, multi-layer perceptron and support vector regression
methods. The dengue cases and meteorological variables were monthly recorded
from $2007$ to $2019$ and from $2005$ to $2019$, respectively. Training the
algorithms, the best performance was obtained by the RF method. With regard to
RF, Ong et al. investigated the dengue transmission in Singapore utilizing data
from the spread of dengue, such as population, entomological and environmental
data \cite{Ong2018}. Their results showed that RF has high accuracy to reproduce
dengue cases, obtaining a correlation coefficient greater than $0.86$. 
They demonstrated that spatial risk of dengue cases can be modeled by means of
RF. For Iquitos (Peru), San Juan (Puerto Rico) and Singapore, Benedum et al.
combined weather data with dengue cases to construct a forecast model
\cite{Benedum2020}. Comparing time series, regression and RF methods, they
verified that the RF method has $21\%$ and $33\%$ of error less than regression
and time series models, respectively, for near short predictions ($4$ up to
$12$ weeks). Mussumeci and Coelho also employed the RF method to forecast dengue
incidence in $790$ cities in Brazil. Concerning the climate data, the authors
considered information about incidence cases in social networks. 

In this work, we employ the RF method \cite{Breiman2001} to forecast dengue
incidence in three cities localized in South America: Natal (Brazil), Iquitos
(Peru) and Barranquilla (Colombia).  In our
simulations, we use the data delayed by up to one week to forecast the new cases. We
also employ three combinations of features: $i$) we took into account
only dengue cases (D); $ii$) we combine dengue and climate data (CD); $iii$) 
we utilize the data of humidity and dengue cases (HD).
{
As an important
new finding, we show that, depending on the city and the training length, the
results can be improved with a given combination of features. For instance, 
for Natal when we consider climate variables the forecasting is not improved. 
For Iquitos, the strategy $ii$ is better for forecasting, while for Barranquilla 
the best strategy is $iii$. Depending on the training length, we find an 
optimal region for each city where the correlation among real and simulated 
data increase.
}

The structure of this work is given by the following order: In Section
\ref{metodos}, we describe the data acquisition processing and the RF method.
Section \ref{times_series} is dedicated to exhibit and extract information of
each time series. Forecasting results are discussed in Section \ref{ml_results}.
Finally, our conclusions are drawn in Section \ref{conclusao}.

%%%%%%%%%%%%%%%%%%%%%%%%%%%%%%%%%
%%%%%%%%%%%%%%%%%%%%%%%%%%%%%%%%%

\section{Methods} \label{metodos}

\subsection{Data acquisition}

In this work, we consider weekly dengue cases and average week climate variables
from three localities: ($i$) Natal (Brazil, elevation $30$m, latitude $-5.81$
and longitude $-35.25$) from the $16$th week of $2016$ until the 52th week of $2019$, 
with time series totalling a length equal to $193$ weeks; ($ii$) Iquitos (Peru,
elevation $106$m, latitude $-3.74$ and longitude $-73.25$) from the 28th week of
$2001$ until the $52$th week of $2012$, whose time series length is equal to $597$
weeks; ($iii$) Barranquilla (Colombia, elevation $18$m, latitude $10.96$ and
longitude $-74.79$) from the 2nd week of $2011$ until the $47$th week of $2016$, with a 
time series length of $307$ weeks. For Natal, we extract the dengue cases from 
Sanchez‐Gendriz et al. \cite{Gendriz2022} and the climate data
(precipitation, relative humidity and air temperature) from National Institute
of Meteorology \cite{Metereologia}. For Iquitos, we obtain both data from the Dengue
Forecasting Project Data Repository \cite{IquitosDados}. 
The climate variables for Iquitos are: minimum and average temperature, 
relative and absolute humidity and precipitation. 
Dengue cases from
Barranquilla are available on Sivigila Portal \cite{Sivigila} and meteorological 
(maxima temperature, relative humidity and precipitation)
in Ref. \cite{Trujillo2018}. 

The data and codes employ in this research 
are available on GitHub \cite{SidneyGithub}.

%%%%%%%%%%%%%%%%%%%%%%%%%%%%%%

\subsection{Data processing and statistical analysis}
When there are many outliers in the data set, a standardization technique helps
in decreasing the error in the results. In this work, we use the Robust Scaler
method, which works by subtracting the median (${\rm med}(X)$) of the data
($X$) and scaling in the interval between the $1$st ($Q_1$) and the $3$rd
($Q_3$) quartiles. The Robust Scaler equation is given by 
\begin{equation}
{\rm RS}(x_i) = \frac{x_i - {\rm med}(X)}{Q_3 - Q_1}.
\end{equation}
In this procedure, the median and interquartile range ($Q_3-Q_1$) are stored and
used on future data as the transformation applied on the forecasting. 

One important characteristic of time series is to know if is stationary 
or non-stationary.  There are some techniques
that are employed to answer this question. In this work, we consider the
augmented Dickey--Fuller test (ADF), which belongs to the unit root test. 
A unit root test verify if the time series is non-stationary (null hypothesis) 
or not (alternative hypothesis). 
The ADF test can be described by 
\begin{equation}
\Delta y_t = \alpha + \beta t + \gamma y_{t-1} + \delta_1 \Delta y_{t-1} + ... + \delta_{p-1} \Delta y_{t-p+1} + \epsilon_t, 
\label{adf}
\end{equation}
where $\alpha$ is a constant, $\beta$ is the time coefficient, $\gamma$ 
is the coefficient that present the root (is the focus of test), 
$p$ is the lag order of the first differences in autoregressive process, 
and $\epsilon$ is a noise term. The parameter that we focus is $\gamma$. 
If $\gamma = 0$, or positive, then we assume the null hypothesis, i.e., 
a non-stationary time series. On the other hand, if $\gamma$ is negative 
we assume the stationarity. In addition to this analyses, we also test 
the null hypothesis by computing the $p$-value. If $p$-value $<0.05$ we 
assume that the time series is stationary. We conducted the ADF test 
by the URCA package in R \cite{urca}.

%%%%%%%%%%%%%%%%%%%%%%%%%%%%%%
\subsection{Random Forest}
\begin{figure}[!ht]
\begin{center}
\includegraphics[scale=0.3]{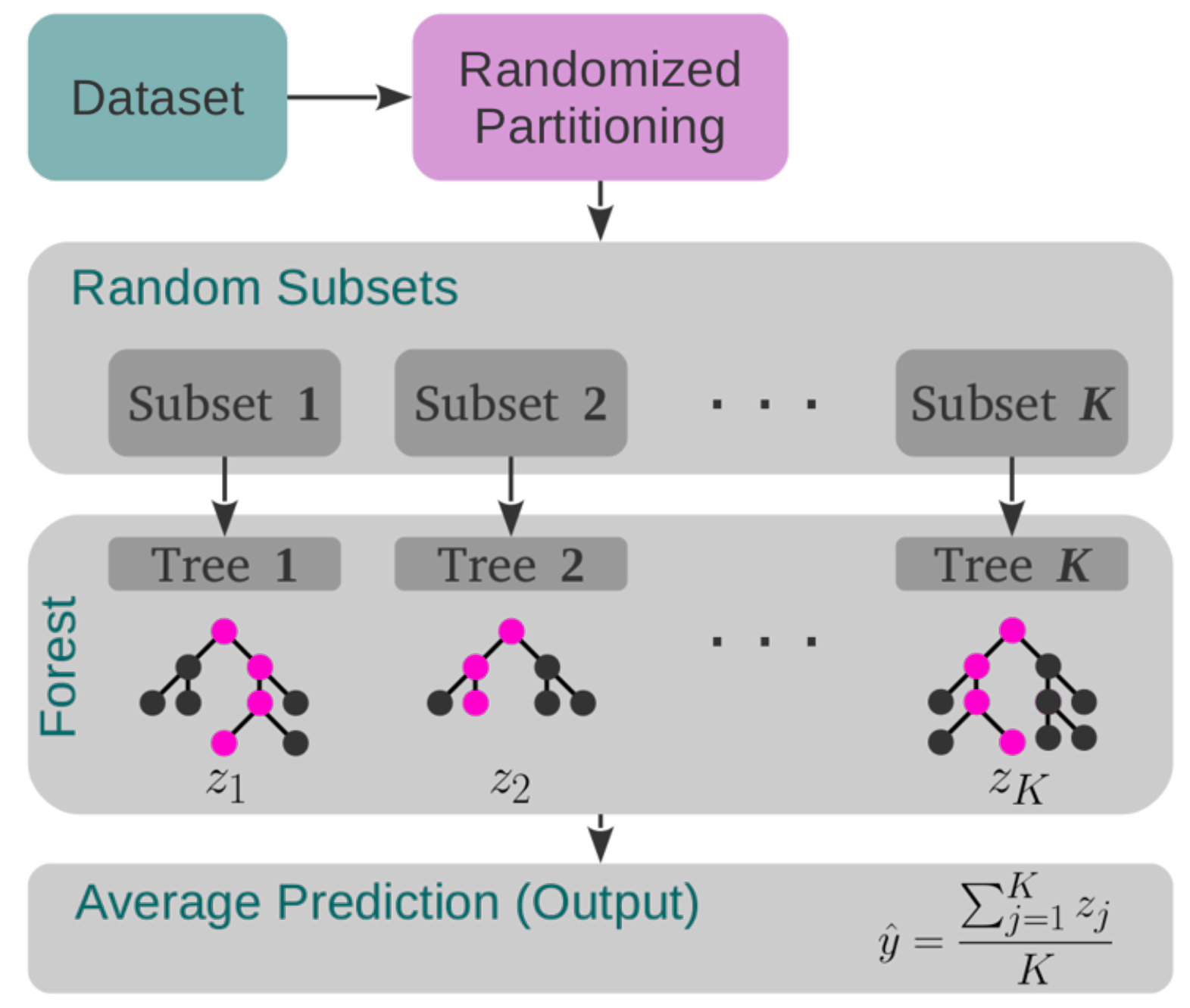}
\caption{Structure of the Random Forest algorithm. $K$ random bootstrap samples
are extracted from the data and an unpruned decision tree is fitted for	each
bootstrap sample. At each node, a small subset of the covariance is randomly
chosen to optimise the split. The forecasting is obtained by averaging the
prediction of all trees.} 
\label{fig1}
\end{center}
\end{figure}

We adopt here the Random Forest (RF) algorithm \cite{Forests}, which is a
supervised ML algorithm based on the ensemble learning method. It is formed by
many decision trees. The ensemble learning techniques are more accurate to made
predictions than individual models. It occurs due to the fact that the ensemble
techniques combine many basic ML algorithms and their predictions. 
Given an initial dataset, the model split it in $K$ random subsets, generating 
different subsets from the original dataset. These final subsets are namely 
terminal or leaf nodes and the intermediate subsets are called internal 
nodes. The prediction of the results in each terminal node is made by 
the average outcome of the training data. In this way, 
the RF algorithm generates predictions or rankings
from a set of multiple decision trees, as schematically represented in Fig.
\ref{fig1}. 

By combining the outputs of the trees, the algorithm provides a consolidated and
more accurate result than the result from basic ML algorithms. The strength of
the algorithm lies in its ability to handle complex data sets and mitigate 
overfitting, making it a valuable tool for various predictive tasks via ML. 

The choice of partitions in each decision tree (Fig. \ref{fig1}) is made 
by considering the minimum Mean Squared Error (MSE)  
\begin{equation}
\label{eq1}
{\rm MSE}_{min} = \min \left\{ \sum_{i\in S_{i}}(y_i - \widehat{y}_i)^{2} \right\},
\end{equation}
where $\widehat{y}$ is the average value of the partition and $y_i$ is the value
of each data point within that partition. Then, it is important to use the 
prune method as a regularization. 
This method makes a balance between the lowest MSE value and its depth according
to
\begin{equation}
{\rm MSE}_{min,\alpha} = \min_{\alpha} \left\{\sum_{i\in S_{i}}(y_i - \widehat{y}_i)^2
+ \alpha T \right\},
\end{equation}
where $\alpha$ is a tuning parameter that is determined by cross validation and
$T$ is the number of terminal nodes \cite{Biau2016}. Our simulations are implemented in Python
and we use sklearn libraries for statically analyses \cite{scikit-learn}.

%%%%%%%%%%%%%%%%%%%%%%%%%%%%%%
\subsection{Error analysis}
In the forecasting range we compute the error among simulated and real points 
by considering the absolute error ($e$), mean absolute error (MAE), and 
the error ($\Delta E$). Given the real time series denoted by $y_i$ and 
the simulated points $x_i$, the absolute error in the $i$--week is defined as 
\begin{equation}
e_i = |y_i - x_i|, 
\end{equation}
the notation $e_{\rm D}$, $e_{\rm CD}$ and $e_{\rm HD}$ means the absolute 
error computed using $x_i$ from strategy D, CD, and HD, respectively. 
Where the index $i$ is omitted by economy of notation. 
Another measure that we use is 
\begin{equation}
{\rm MAE} = \frac{1}{n} \sum_{i=1}^{n} |y_i - x_i|,
\end{equation}
where $n$ is the number of weeks. 

Another important question to study is whether 
our model leads to overestimating or underestimating cases. To extract this
information, we compute 
\begin{equation}
\Delta E_i = y_i - x_i,
\end{equation}
then, when $\Delta E_i < 0$ the model overestimating, and $\Delta E_i>0$ 
the forecasting is underestimating.
%%%%%%%%%%%%%%%%%%%%%%%%%%%%%%%%%
%%%%%%%%%%%%%%%%%%%%%%%%%%%%%%%%%

\section{Time series} \label{times_series}

\subsection{Natal (Brazil)}

Time series for Natal are displayed in Fig. \ref{fig2}, where the panel (a)
shows the dengue cases ($\times 10^2$), (b) precipitation (mm) and (c) relative
humidity (\%). The panel (d) exhibits the minimum (blue line), maximum (red
line) and average temperature (black line) in $^{\rm o}$C. The bottom $x$-axis display
the year and its relative epidemiological week, while the upper $x$-axis shows the
number of week for the whole time series. We compute the power spectra density 
using the package PDSR implemented in R \cite{pdsr_r} for all the variables. 
{Our results show that the period is approximately 1 year for all variables. 
Specifically, for the dengue cases we obtain a period equal to $53 \pm 2$ weeks. 
Before employing the ML algorithm, we check the
stationary of the data by means of the ADF method. The result of the test is
displayed in Table \ref{Table1}. In the ADF test present in this paper, 
we set the $p$-value equal to zero when the algorithm returns a value 
less than $10^{-6}$.  As the $p$-value for the dengue cases is greater 
than 0.05 and $\gamma$  is close to zero, we conclude that this 
time series is non-stationary.  
Figure \ref{fig2}(a) exhibits the data
during the four analysed years. From the 16th week of 2016 until the end of this
year, 3422 cases were reported. In 2017, 2018 and 2019, there were 4747,
15178 and 17197 cases reported, respectively. In 2018 and 2019, there were more cases
than expected, then in these years  outbreaks occurred. Considering the time
range of the first outbreak from the 10th until the 37th week of 2018 (where more than
200 cases occur every week), there were 12442 reported cases. Using the same
criteria, the second outbreak exists between the 12th until the 38th week of 2019,
where 14687 cases were reported. Just these two outbreaks were responsible for
27129 infections. 

\begin{table}[!ht]
\caption{Test of stationarity using ADF Test for the Natal features.}
\begin{center}
\scalebox{0.9}{
\begin{tabular}{|c|c|c|} 
\hline  
\small Feature & \small $p$-value & \small $\gamma$\\
\hline
Cases          &        $0.088$    & $-0.053 \pm 0.024$        \\
\hline 
Precipitation  &        $0$        & $-0.803 \pm 0.070$            \\
\hline 
\small Rel. Humidity &  $0$        & $-0.421 \pm 0.058$       \\
\hline 
Min. Temp.     &        $8 \times {10^{-4}}$ & $-0.150 \pm 0.040$            \\
\hline 
Max. Temp.     &        $4 \times 10^{-5}$ & $-0.202 \pm 0.044$          \\
\hline 
Avg. Temp.     &        $9 \times {10^{-4}}$ & $-0.143 \pm 0.037$        \\
\hline 
\end{tabular}}
\end{center}
\label{Table1}
\end{table}

\begin{figure}[!ht]
\begin{center}
\includegraphics[scale=0.8]{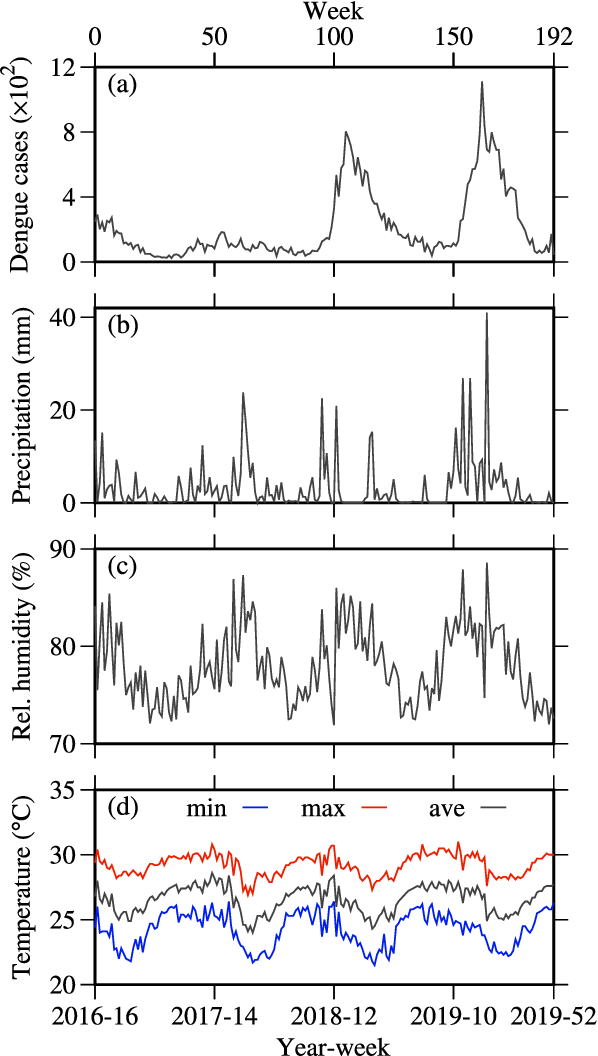}
\caption{Time series from Natal for (a) dengue cases, (b) precipitation, (c)
relative humidity and (d) air temperature.}
\label{fig2}
\end{center}
\end{figure}

\subsection{Iquitos (Peru)}

The time series for the data from Iquitos are displayed in Fig. \ref{fig3}, 
where the panel (a) is for dengue cases ($\times 10$), (b) is for precipitation
(mm), (c) is for absolute humidity (\%), (d) for relative humidity (\%) and 
(e) is for minimum (blue line) and average temperature $^{\rm o}$C (black line). The
considered range for these data varies from the 28th week of 2001 until the 52th week of
2012, totalling 597 weeks (marked in the top $x$-axis). 
By the power spectra density analyze, our results shows a period of 
1 year for all the variables. More precisely, the cases has a period 
equal to $55 \pm 3$ weeks. The ADF test shows that the times
series related to Iquitos are stationary, as observed in Table \ref{Table2}. 
The associated $p$-values in Table \ref{Table2} are setted equal zero because 
our test report values less than $10^{-6}$. 
In
terms of reported cases, Iquitos received  274, 490, 171, 715, 451, 256, 562,
694, 296, 585, 95 and 501 reports during 2001-2012, respectively. Along this
time series, we observe the presence of 9 outbreaks along these 12 years. Two of
them call  attention due to the high amplitude. The first one occurs between the
22th and the 27th weeks of 2004 and the second between the 25th and the 35th weeks of 2010,
occurring a total of infection equal to 265 and 413, respectively.  

\begin{table}[!ht]
\caption{Test of stationarity using ADF Test for the Iquitos features.}
\begin{center}
\scalebox{0.9}{
\begin{tabular}{|c|c|c|} 
\hline  
\small Feature & \small $p$-value & $\gamma$ \\
\hline
Cases          &        $0$ & $-0.221 \pm  0.024$  \\
\hline 
Precipitation  &        $0$ & $-0.855 \pm  0.040$   \\
\hline 
\small Spec. Humidity&  $0$ & $-0.414 \pm 0.033$   \\
\hline 
\small Rel. Humidity&   $0$ & $-0.322 \pm 0.030$   \\
\hline 
Min. Temp.&             $0$ & $ -0.300 \pm 0.030$  \\
\hline 
Avg. Temp.&             $0$ & $-0.432 \pm  0.033$   \\
\hline 
\end{tabular}}
\end{center}
\label{Table2}
\end{table}

\begin{figure}[!ht]
\begin{center}
\includegraphics[scale=0.8]{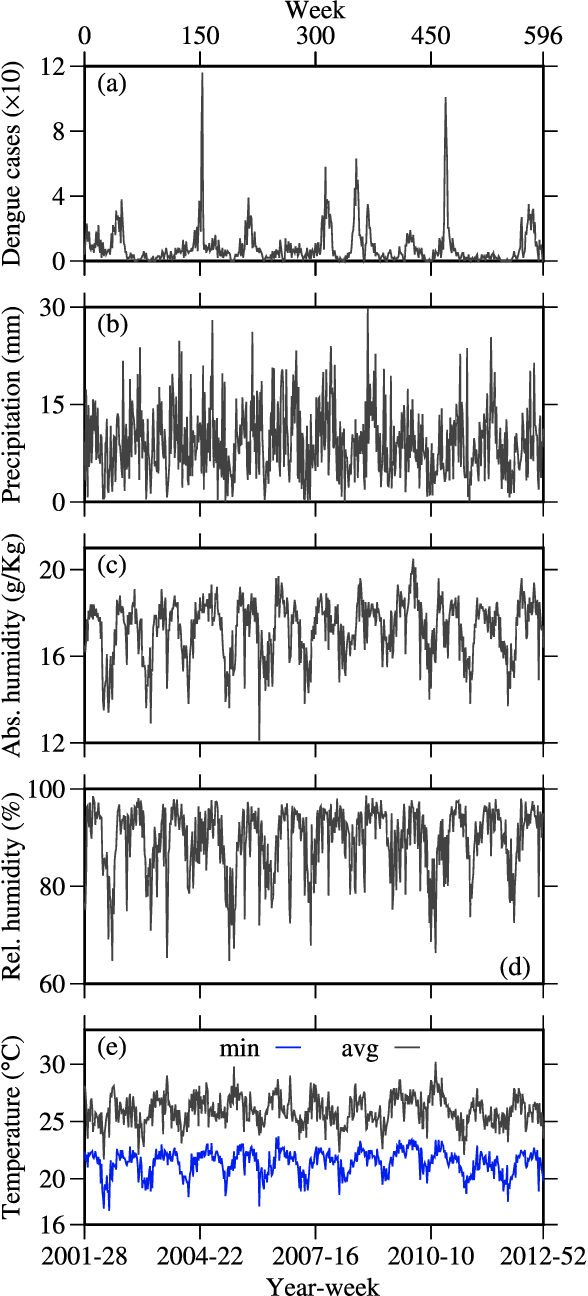}
\caption{Time series from Iquitos for (a) dengue cases, (b) precipitation, (c)
specific humidity, (d) relative humidity and (e) air temperature.}
\label{fig3}
\end{center}
\end{figure}

%%%%%%%%%%%%%%%%%%%%%

\subsection{Barranquilla (Colombia)}

Figure \ref{fig4} displays the time series for Barranquilla. The panels (a), 
(b), (c) and (d) exhibit the time evolution of the dengue cases ($\times 10^2$),
maximum air temperature $^{\rm o}$C, precipitation (mm) and relative humidity (\%),
respectively. The data for Barranquilla start in the first week of 2011 and ends
in the 47th week of 2016 (bottom $x$-axis), totalling 307 weeks (top $x$-axis). 
From the
spectral analyzes, the period associated with the climate variables is 1 year. 
The period for dengue 
cases in Barranquilla correspond to $52 \pm 4$ weeks. 
The time series from Barraquilla are stationary,
as observed in Table \ref{Table3}. Again, $p$-values less than $10^{-6}$ 
are considered equal to zero.
If we consider an outbreak when more than 50
cases are reported, we see 3 outbreaks in Fig. \ref{fig4}(a). The first one
starts in the 8th week of 2013 and ends in the 52th week of the same year, having 2425
reported cases. The second one starts in the 39th and finishes  in the 53th week of 2014
where 1210 cases were reported. The last one occurs from the 43th week of 2015 and 
finishes in the 3rd week of 2016, with 979 cases. Due to the huge outbreaks in 2013 
and 2014, they are the years with more reported cases. For instance, 655, 946,
1364 and 617 cases were reported in 2011, 2012, 2015 and 2016. However, the data
show 2748 and 2737 cases in 2013 and 2014, respectively.  

\begin{table}[!ht]
\caption{Test of stationarity using ADF Test for the Barranquilla features.}
\begin{center}
\scalebox{0.90}{
\begin{tabular}{|c|c|c|} 
\hline  
\small Feature & \small $p$-value & $\gamma$ \\
\hline
Cases          & $0.0001$ & $-0.116  \pm 0.027$  \\
\hline 
Max. Temp.     & $0$      & $-0.293  \pm 0.040$   \\
\hline 
Precipitation  & $0$      & $-0.696  \pm 0.054$   \\
\hline 
\small Rel. Humidity& $0$ & $-0.271 \pm  0.040$   \\
\hline 
\end{tabular}}
\end{center}
\label{Table3}
\end{table}

\begin{figure}[!ht]
\begin{center}
\includegraphics[scale=0.8]{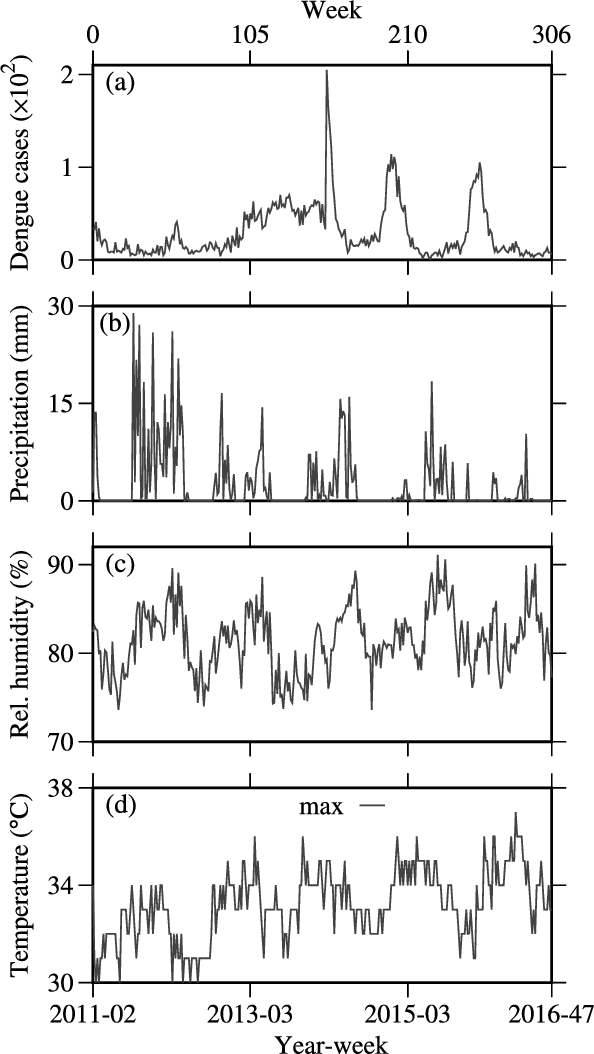}
\caption{Time series from Barranquilla for (a) dengue cases, (b) precipitation, 
(c) relative humidity and (d) maximum air temperature.}
\label{fig4}
\end{center}
\end{figure}

%%%%%%%%%%%%%%%%%%%%%%%%%%%%%%%%%
%%%%%%%%%%%%%%%%%%%%%%%%%%%%%%%%%

\section{ML based forecasting} \label{ml_results}

After analysing the general characteristics of the time series, we train the RF
algorithm. We consider three approaches to compare the predictive capacity of
the algorithm: climate data and dengue cases (CD) in the $i$-th week to predict
dengue cases in the $(i+1)$-th week; dengue cases (D) from the $i$-th week to
predict disease cases in the $(i+1)$-th week; humidity and dengue cases (HD)
from the $i$-th week to predict the cases in the $(i+1)$-th week. The data are separated
into the training and testing stages according to the length of the time series.

\subsection{Natal forecasting}

Firstly, we employ the algorithm in Natal with training a length equal to 144
weeks and the 49 remaining weeks. In our algorithm, we consider the number of
trees, maximum features and depth equals to 1000, 0.5 and 100, respectively. The
criteria to choose partitions is Friedman MSE \cite{Biau2016}. For CD, we consider previous
cases, minimum, maximum and average temperature, humidity and precipitation. For
each feature, the algorithm gives the following importance: 0.67, 0.04, 0.03,
0.03, 0.12 and 0.08. The dengue cases and humidity are more often used by the algorithm. 
Figure \ref{fig5}(a) displays the dengue cases (black points) and our simulated
results (colored lines). The training region shows the red (CD approach) and
the dotted blue (D approach) line, while the test range (grey background) exhibits
the magenta (CD) and green (D) lines. As expected, in the train region, we
observe a good accordance among the points generated by the ML algorithm and the
real data. In the test region, magnified in the panel (b), the error increases
when compared with the training. In the panel (b), the new curves exhibit the
absolute error ($e$) among the real and simulated points. The light magenta
colour, denoted by $e_{\rm CD}$, is the absolute error that is associated with
the CD approach, while the light green ($e_{\rm D}$) with $D$ approach. Our results show
that the error increases in the peak region, in both approaches. The mean absolute
error (MAE) for CD in the test range is 97.92 and the correlation coefficient
($r$) is 0.90, while for D approach these values change to 62.27 and 0.94,
respectively. Considering that the peak starts in the 156th week and ends in the 173th
week, the MAE before the peak is 49.24 and 42.08 for the CD and D approach.
During the peak, the MAE is 166.67 and 88.34 for CD and D, respectively. After
the peak the MAE is 35.44 and 52.22 for CD and D. During and before the peak, D
performs better than the CD approach. This scenario changes after the peak.
During the test region, the maximum value of reported cases is 1113 and the MAE
in peak range represents approximately 14.9\% and 7.9\% of this value. Outside
this range, the MAE is around 4\%.  The
result for $\Delta E_i$ in the testing range, is displayed in Fig. \ref{fig5}(c) by the blue
line for CD and the black line for D.  Before the
peak, we have mostly underestimating values. After the peak, we identify a mix.
This overestimation occurs due to the fact that the algorithm learns this
huge peak and this information remains in its memory.

\begin{figure}[!ht]
\begin{center}
\includegraphics[scale=0.8]{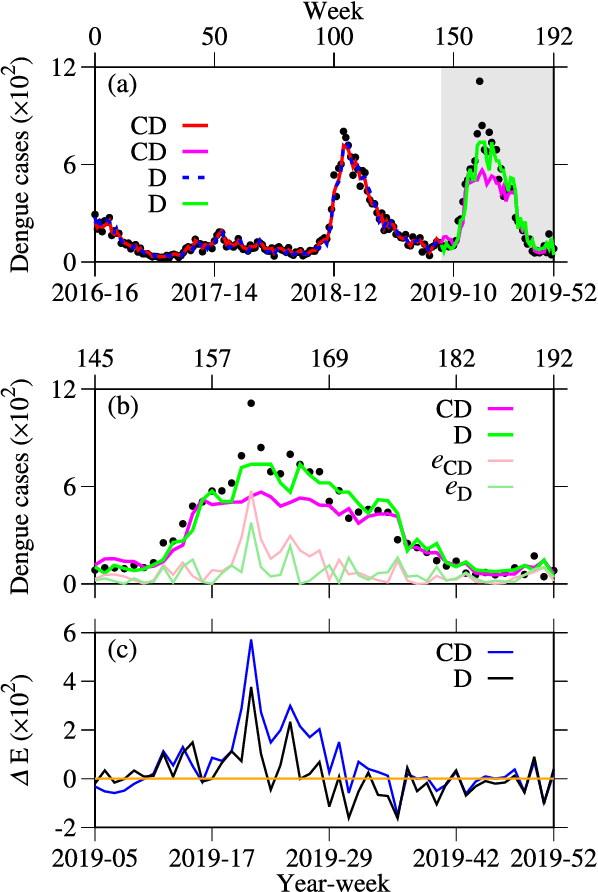}
\caption{(a) Dengue cases (black points) and ML forecast results for Natal. The
red and dotted blue lines show the training range for CD and D approach,
respectively. The test range is highlighted by the gray background, where the
magenta and green lines display the results by considering the CD and D
approaches, respectively. (b) Magnification of testing region. The light magenta
and blue curves display the absolute error $e$ among real data and CD and D
approach, respectively. (c) Error among real and simulated data.}
\label{fig5}
\end{center}
\end{figure}

Figure \ref{fig6} displays the comparison between each approach, i.e., CD (blue
line), D (dotted black line) and HD (dotted red line), as a function of the
training length. The panel (a) exhibits the MAE and the panel (b) shows the $r$
computed in the remaining test region. The main results are: when we
train our algorithm for a few weeks, i.e., less than 55\% of the time series, the
RF produces considerable error in the forecast. Increasing the training length, 
the RF starts to perform better in the forecast. If we consider more than 90\%
of our time series for training, the error increases and the correlation
decreases. After 90\% of training, just a few weeks remain for forecast, then we
do not have enough weeks to obtain a reasonable statistic. Utilizing between
60\% and 89\% of the time series for training, we observe that the MAE decreases
and $r$ increases. In this range, a better performance is obtained for the D
approach. For a training length inferior to 55\%, the CD and HD perform better
than D approach.

\begin{figure}[!ht]
\begin{center}
\includegraphics[scale=0.8]{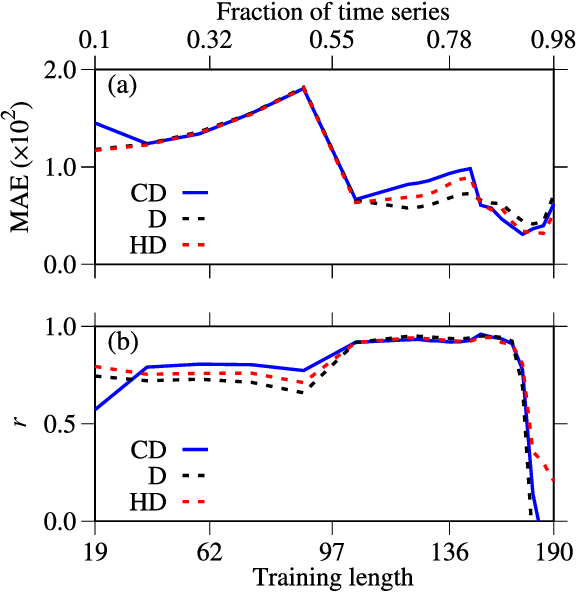}
\caption{Comparison between the CD, D and HD approaches as a function of 
the training length for Natal data. (a) Mean absolute error (MAE) and (b) correlation coefficient ($r$) as
function of the length training. The blue line displays the
results for CD approach, while the dotted black and red lines for D and HD,
respectively.}
\label{fig6}
\end{center}
\end{figure}

\subsection{Iquitos forecasting}

The second analyzed city is Iquitos. There we consider 333 weeks of
training and 263 of forecasting.  We choose this training length 
(56\% of the whole time series) due the fact that those 333 weeks
separated two, apparently, symmetric outbreaks. The result is exhibited in Fig.
\ref{fig7}(a), where the red and dotted blue lines are related to training and
the magenta and green lines to the test region for the CD and D approach, respectively.
The black points exhibit the dengue cases ($\times 10$). The hyper parameters
are the number of trees equal to 200, maximum features equal to 0.6, depth equal to
50 and criteria to choose partitions equal to the Absolute Error. In the CD
approach, we consider the following features: Cases, minimum and average temperature,
relative and absolute humidity and precipitation. The algorithm weights these
features as 0.39, 0.11, 0.10, 0.13, 0.13 and 0.13, respectively, in the learning
process. For this, humidity and precipitation are mostly used. In
addition, we consider a very long horizon forecast (up to 263 weeks). We obtain one MAE $= 4.42$ 
and $r = 0.83$ in the test range for the CD situation. For D approach, these
values change to 4.02 and 0.81, showing that CD performs better than D in this
case. The first outbreak in the testing range has a symmetric series of outbreaks.
Due to this reason, we choose this training length. Doing this consideration,
we observe that RF performs a good forecast, as displayed in the magnification
in Fig. \ref{fig7}(b). As in the previous subsection, the algorithm also fails in
predicting the highest peaks. The better performance in this situations seems to be in
the D approach, however, this is not true. By looking for the MAE along the time
series, as exhibited in Fig. \ref{fig7}(c), we verify that the error associated
with the CD approach (light magenta line, denoted by $e_{\rm CD}$) is less in the outbreaks when compared
with the errors associated with the D situation (light green line, denoted by $e_{\rm D}$). Also, after the
huge peak localized at the 467th week, the D approach exhibits an overestimation of
cases (panel (e)) higher than the CD approach (panel (d)).

\begin{figure}[!ht]
\begin{center}
\includegraphics[scale=0.8]{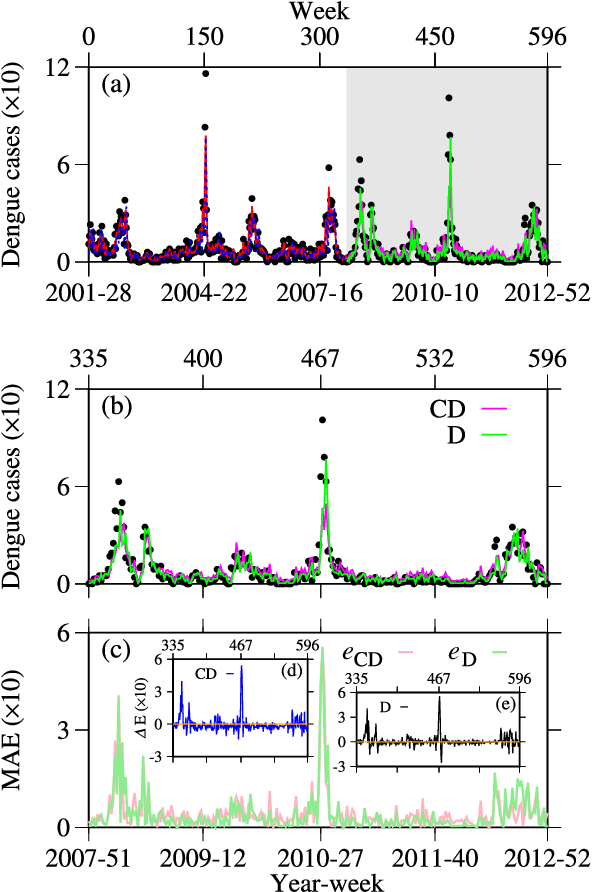}
\caption{(a) Iquitos dengue cases (black points) and ML forecast results. The
red and dotted blue lines show the training range for the CD and D approaches,
respectively. The test range is highlighted by the gray background, where the
magenta and green lines display the results by considering the CD and D
approaches, respectively. (b) Magnification of testing region. (c) MAE among
real and simulated data, where the light magenta line is associated with CD and
the light green with D. (d) Absolute error for the CD approach. (e) Absolute
error for the D approach.}
\label{fig7}
\end{center}
\end{figure}

MAE and $r$ can be improved if we train our algorithm for longer time (Fig. \ref{fig8}). 
If we take less than 55\% of the time series to train
the algorithm, MAE increases and $r$ decreases. In addition, for certain values,
CD  (blue line) performs worse than D (black dotted line) or HD (red dotted
line) approaches. For values after week 331 (55\% of time series) of
training, the results start to improve, in the sense of obtaining a minimum MAE
and maximum $r$, until a threshold value that occurs when we use 90\% of the 
time series for training. At this point, MAE and $r$ are respectively equal to 
3.57 and 0.88 for CD, 4.06 and 0.77 for D, and 3.53 and 0.85 for HD. We see that
CD and HD perform better than D. After 90\%, MAE increases and $r$ decreases. 
Besides, there are enough weeks after this limit (less than 60). In this given
range, the algorithm does not perform very well in forecasting. The range captures
just the last outbreak of the time series, where the points oscillate very
irregularly. 

\begin{figure}[!ht]
\begin{center}
\includegraphics[scale=0.8]{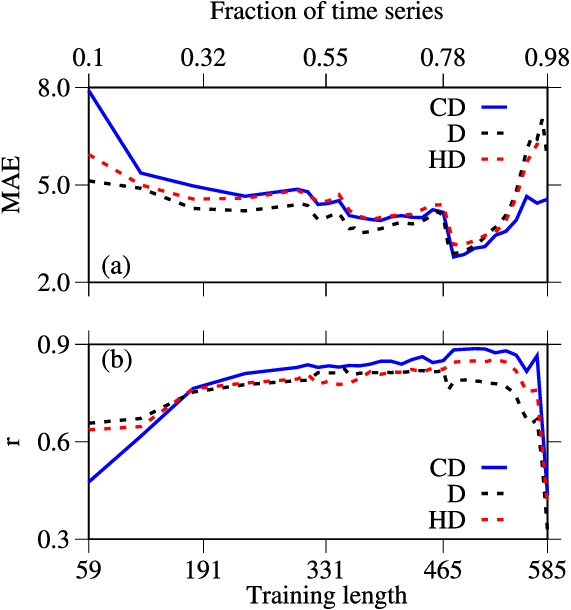}
\caption{Comparison between the CD, D and HD approaches as a function of the training length for Iquitos data. 
(a) Mean absolute error (MAE) and (b) correlation coefficient ($r$) as
a function of length training. The blue line displays the
results for CD approach, while the dotted black and red lines for D and HD,
respectively.}
\label{fig8}
\end{center}
\end{figure}

\subsection{Barranquilla forecasting}

The last analyzed city is Barranquilla. For this city our simulations show  
better results using the HD approach then the D approach. Figure \ref{fig9}(a) 
displays the real data for dengue cases ($\times 10$) by the black points, 
red and blue lines the training range, and magenta and green the testing 
curves for CD and HD approaches, respectively. In our simulation, we take  
233 weeks for training, corresponding to 76\% of the time series length and
remaining 74 weeks of forecasting. We use this training length to observe the
forecast of the last outbreak. For the hyper parameters the number of trees, 
maximum features, depth and criteria to choose partitions, we utilize 1500, 0.2,
200 and Poisson. In the CD approach, the ML uses 0.65, 0.10, 0.16 and 0.10 of
the features cases, maximum temperature, relative humidity and precipitation, 
respectively. In the HD features, the algorithm provides 0.82 and 0.17 of
importance to cases and relative humidity, respectively. For the first approach,
we obtain 7.81 for MAE and 0.92 for $r$, while for HD, the values are 6.22 and
0.94. The testing range is amplified in Fig. \ref{fig9}(b), where the light
magenta and green curves show $e_{\rm CD}$ and $e_{\rm D}$. It is important to note 
that the fitting of the peak performs better than in the Natal case. Considering
the peak region defined between the 250th until the 262th week, the CD approach
returns one MAE equal to 15.90 and HD equal to 13.16. As also observed in Natal,
the approach without all considered climate variables performs better in the
peak range. From the point where we started the training until 249 the MAE for
CD and HD are, respectively, 3.63 and 5.00. After 262 weeks, these values are
7.09 and 4.69. The CD approach performs better before the peak. Another
characteristic that emerges after 262 weeks is that the ML overestimates the
forecast values, as observed by the results in Fig. \ref{fig9}(c).

\begin{figure}[!ht]
\begin{center}
\includegraphics[scale=0.8]{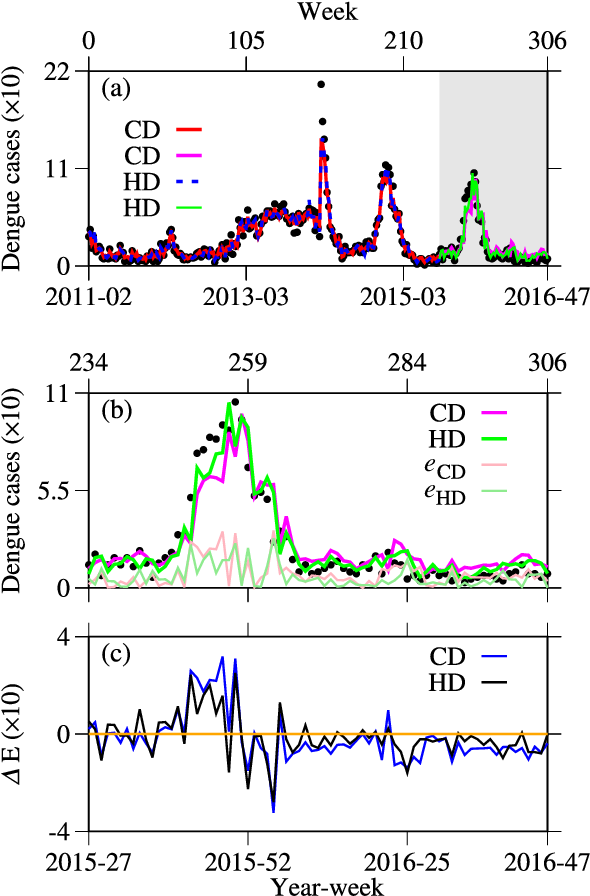}
\caption{(a) Barranquilla dengue cases (black points) and ML forecast results.
The red and dotted blue lines show the training range for CD and HD approaches,
respectively. The test range is highlighted by the gray background, where the
magenta and green lines display the results by considering the CD and HD
approaches, respectively. (b) Magnification of testing region. (c) Error among
real and simulated data.}
\label{fig9}
\end{center}
\end{figure}

A comparison among the approaches CD (blue line), D (black dotted line), 
and HD (red dotted line) is displayed in Fig. \ref{fig10}, where panel (a) 
exhibits MAE and panel (b) shows $r$ as a function of training length. As 
observed in the previous results, the algorithm performs better in the testing
region, when we consider more than 55\% of the time series as training. However,
in this case, if we use more than 85\% of the time series as training, the error
increases and the correlation decreases. It is important to note that if we
take the approach D and more than 96\% of the time series as training, the
correlation becomes negative. The algorithm forecast increases in the cases when
the data show a decay. Comparing the methods, D performs better just when we use
less than 40\% of the time series as training. But in the range where we get 
a good correlation and small error, i.e., between 55\% and 85\% the best
strategy is consider the HD approach. 

\begin{figure}[!ht]
\begin{center}
\includegraphics[scale=0.8]{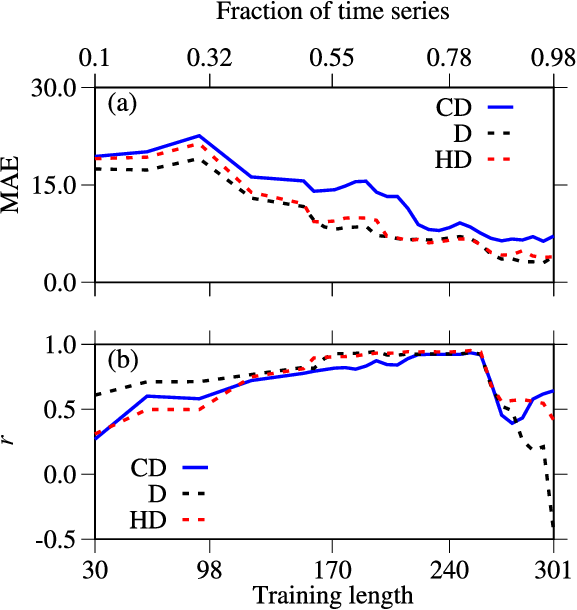}
\caption{Comparison between the CD, D and HD approaches as a function of the training length for Barranquilla data. 
(a) Mean absolute error (MAE) and (b) correlation coefficient ($r$) as
a function of length training. The blue line displays the
results for the CD approach, while the dotted black and red lines for D and HD
approaches, respectively.}
\label{fig10}
\end{center}
\end{figure}

%%%%%%%%%%%%%%%%%%%%%%%%%%%%%%%%%
%%%%%%%%%%%%%%%%%%%%%%%%%%%%%%%%%

\section{Conclusion} \label{conclusao}

In this work, we employ Random Forest (RF) machine learning (ML) technique to forecast
dengue infections based on previous cases and meteorological variables. We test
our approach for data from Natal (Brazil), Iquitos (Peru) and Barranquilla (Colombia). In
our simulations, we use three approaches: where we consider only dengue cases 
(D); a combination of climate variables and dengue cases (CD); the combination of
humidity and dengue cases (HD). We use the features from the $i$-week to forecast the
cases in ($i+1$)-week. We also test different delays in the three cities, however,
we do not obtain improvement. For each city, we use a set of climate variables. 
For Natal, we utilize the average, minimum, maximum temperature, precipitation
and relative humidity. We find that the algorithm uses more  humidity
and the dengue cases in the forecasting process. Nevertheless, for Natal, we get an
improvement including the dengue cases in the prediction only. For Iquitos, we
consider the relative humidity, minimum and average temperature, specific
humidity and precipitation. In this case, the humidity and precipitation are
used with more weight by the ML technique. The best forecasting performance for Iquitos
occurs, when climate and dengue cases are included. For
Barranquilla, we utilize the maximum air temperature, relative humidity and
precipitation. Then the most important variable considered by the
ML is the relative humidity. For the data from Barranquilla, our results show a
better performance, when there is the combination of dengue and humidity data.

A common characteristic that emerges is that the forecast generated by the RF
algorithm exhibits a higher error in the peak regions. Moreover, exploring the
effects of training length, we find that the algorithm performs
better when we use more than 55\% of the time series and less than 90\%. The
optimal region for Natal occurs when we include more than 64\% and less than
80\% of the time series for training. In this range, the error decreases and
the correlation increases. The best performance is generated by the D method.
For this approach, we verify correlations varying between 0.917 and 0.949, and
MAE between 57.783 and 71.768. The optimal range for Iquitos occurs considering
between 79\% and 88\% of the time series as training length. The method that
performs better is CD, having a minimum $r$ equal to 0.850 and maximum equal to
0.887 while MAE oscillates between 2.780 and 4.156. For Barranquilla, the
optimal range occurs between 72\% and 82\% of length training. The method that
performs better in this range is HD with minimum $r$ equal to 0.942 and maximum
equal to 0.953 while the minimum and maximum MAE are equals to 6.085 and 6.669,
respectively. The ML technique reproduces many outbreaks and after that there is no
outbreak. Then, the algorithm overestimated some cases, as we observed in
Barranquilla. Another situation is that there are only few points in order to do
statistical analysis, as in Natal. For Iquitos, the high values of training
length select a very noisy region in the time series. The region is
characterized by one outbreak and many fluctuations that are not predicted by RF.

Besides the dengue vector needs proper climate conditions. Our work exhibits that
there is not a preferable climate variable in the forecasting process by using
the RF method. On the other hand, climate variables can increase the error and
decrease the correlation in some situations. In addition, also as a novelty, we
find that the humidity can be more relevant than other climate variables in the
forecasting process. Our results  were also tested using the classic algorithms CNN 1D,
LSTM, ARIMA and SARIMA regression and also we normalize the humidity data
removing the annual component. However, the RF algorithm displays a better
result for our goal and the modification in the humidity did not improve our
results. 

Our work shows that the RF method is useful for the forecasting of new dengue
cases based on meteorological characteristics. This methodology can be employed
by public health organization to forecast and study control measures, because RF is fast,
efficient, robust, and exhibits a high correlation and low mean absolute error
in the forecasting. For future works, we plan to explore other characteristics,
such as mosquito eggs, and contamination rates to better predict new
cases.

%%%%%%%%%%%%%%%%%%%%%%%%%%%%%
%%%%%%%%%%%%%%%%%%%%%%%%%%%%%

\section*{Acknowledgements}
This work was possible with partial financial support from the following 
Brazilian government agencies: CNPq, CAPES, Funda\-\c c\~ao A\-rauc\'aria 
and S\~ao Paulo Research Fo\-undation (FAPESP 2018/03211-6, 2022/13761-9, 2020/04624-2, 2023/12863-5.). 
E.C.G. received partial financial support from Coordena\c c\~ao de
Aperfei\c coamento de Pessoal de N\'ivel Superior - Brasil (CAPES) - Finance
Code \newline{88881.846051/2023-01}. We thank 105 Group Science
\newline(www.105groupscience.com).

\section*{Data availability}
The datasets generated during and/or analyzed during the current study are available 
on GitHub \cite{SidneyGithub} and also can be request for the corresponding author.

%%%%%%%%%%%%%%%%%%%%%%%%%%%%%
%%%%%%%%%%%%%%%%%%%%%%%%%%%%%


\begin{thebibliography}{}
\bibitem{Lancet}
Dengue emergency in the Americas: time for a new continental eradication plan. 
{\it The Lancet Regional Health - Americas}. Available on: https://doi.org/10.1016/j.lana.2023.100539 
Accessed on: 08 March 2024.

\bibitem{PAHO}
Epidemiological Alert - Increase in dengue cases in the Region of the Americas - 16 February 2024. 
{\it Pan American Health Organization}. Available on: https://www.paho.org/en/documents/epidemiological-alert-increase-dengue-cases-region-americas-16-february-2024 
Accessed on: 08 March 2024.

\bibitem{Bhatt2013}
S. Bhatt, P.W. Gething, O.J. Brady, J.P. Messina, A.W. Farlow, 
C.L. Moyes, J.M. Drake, J.S. Brownstein, A.G. Hoen, 
O. Sankoh, M.F. Myers, D.B. George, T. Jaenisch, 
G.R.W. Wint, C.P. Simmons, T.W. Scott, J.J. Farrar, S.I. Hay, 
Nature {\bf 496} 504-507 (2013).

\bibitem{WHO2023a}
World Health Organization. Dengue. Geneva: WHO, 2023.

\bibitem{Rosen1983}
L. Rosen, D.A. Shroyer, R.B. Tesh, J.E. Freier, J.C. Lien 
Amer. J. of Trop. Med. and Hyg. {\bf 32} 1108-1119 (1983).

\bibitem{WHO}
World Health Organization. Dengue and dengue hemorrhagic
fever. Geneva: WHO, 2023.

\bibitem{Messina2014}
J.P. Messina, O.J. Brady, T.W. Scott, C. Zou, D.M. Pigott, K.A. Duda, 
S. Bhatt, L. Katzelnick, R.E. Howes, K.E. Battle, C.P. Simmons, 
S.I. Hay, Trends in Micr. {\bf 22} 138-146 (2014).

\bibitem{Simmons2012}
C.P. Simmons, J.J. Farrar, N. van Vinh Chau, B. Wills, 
New Eng. J. of Med. {\bf 366} 1423-1432 (2012).

\bibitem{Halstead2003}
S.B. Halstead, Adv. in Virus Res. {\bf 60} 421-467 (2003).

\bibitem{Shepard2016}
D.S. Shepard, E.A. Undurraga, Y.A. Halasa, J.D. Stanaway, 
The Lan. Inf. Dis. {\bf 16} 935-941 (2016).

\bibitem{ECDPC}
European Centre for Disease Prevention and Control. 
{Dengue world-wide overview}. An agency of the European Union, 2023.

\bibitem{Morin2013}
C.W. Morin, A.C. Comrie, K. Ernst, Env. Health Per. {\bf 121} 11-12 (2013).

\bibitem{Watts1987} 
D.M. Watts, D.S. Burke, B.A. Harrison, R.E. Whitmire, A. Nisalak, 
The Amer. J. of Trop. Med. and Hyg. {\bf 36} 143-152 (1987).

\bibitem{Reinhold2018}
J.M. Reinhold, C.R. Lazzari, C. Lahondere, 
Ins. {\bf 9} 158 (2018).

\bibitem{Garin2000} 
A. de Gar\'in , R.A. Bejar\'an, A.E. Carbajo, S. de Casas, N.J. Schweigmann, 
Int. J. of Biom. {\bf 44} 148-156 (2000).

\bibitem{Eisen2014}
L. Eisen, A.J. Monaghan, S. Lozano-Fuentes, D.F. Steinhoff, M.H. Hayden, 
P.E. Bieringer, J. of Med. Ent. {\bf 51} 496-516 (2014).


\bibitem{Alto2001} 
B.W. Alto, S.A. Juliano, J. of Med. Ent. {\bf 38} 646-656 (2001).

\bibitem{Lega2017}
J. Lega, H.E. Brown, B. Barrera, J. of Med. Ent. {\bf 54} 1375-1384 (2017).

\bibitem{Kraemer2019} 
M.U.G. Kraemer, R.C. Reiner, O.J. Brady, J.P. Messina, M. Gilbert, 
D.M. Pigott, D. Yi, K. Johnson, L. Earl, L.B. Marczak, S. Shirude, 
N.D. Weaver, D. Bisanzio, T.A. Perkins, S. Lai, X. Lu, P. Jones, 
G.E. Coelho, R.G. Carvalho, W.V. Bortel, C. Marsboom, G. Hendrick, 
F. Schaffner, C.G. Moore, H.H. Nax, L. Bengtsson, E. Wetter, 
A.J. Tatem, J.S. Brownstein, D.L. Smith, L. Lambrechts, 
S. Cauchemez, C. Linard, N.R. Faria, O.G. Pybus, 
T.W. Scott, Q. Liu, H. Yu, G.R.W. Wint, S.I. Hay, Nat. Micr. {\bf 4} 854-863 (2019).

\bibitem{Xavier2021}
L.L. Xavier, N.A. Hon\'orio, J.F.M. Pessanha, P.C. Peiter, Plos One 
{\bf 16} e0251403 (2021).

\bibitem{Lowe2016} 
R. Lowe, C.A. Coelho, C. Barcellos, M.S. Carvalho, R.D.C. Cat\~ao, 
G.E. Coelho, W.M. Ramalho, T.C. Bailey, D.B. Stephenson, 
X. Rod\'o, eLife {\bf 5} e11285 (2016).

\bibitem{Oki2012}  
M. Oki, T. Yamamoto, Plos one {\bf 7} e48258 (2012).

\bibitem{Paul2015} 
B. Paul, W.L. Tham, Health {\bf 7} 672-678 (2015).

\bibitem{Johansson2009}  
M.A. Johansson, F. Dominici, G.E. Glass, Plos Neg. Trop. Dis. 
{\bf 32} e382 (2009).

\bibitem{Johansson2016} 
M.A. Johansson, N.H. Reich, A. Hota, Sci. Rep. {\bf 6} 33707 (2016).

\bibitem{Yuan2020}
H.-Y. Yuan, J. Liang, P.-S. Lin, K. Sucipto, M.M. Tsegaye, 
T.-H. Wen, S. Pfeiffer, D. Pfeiffer, Sci. Rep. {\bf 10} 4297 (2020).

\bibitem{Carvajal2018}
T.M. Carvajal, K.M. Viacrusis, L.F.T. Hernandez, H.T. Ho, D.M. Amalin, 
K. Watanabe, BMC Inf. Dis. {\bf 18} 183 (2018).

\bibitem{Appice2020}
A. Appice, Y.R. Gel, I. Iliev, V. Lyubchich, D. Malerba, 
IEEE Acc. {\bf 8} 52713–52725 (2020).

\bibitem{Guo2017}
P. Guo, T. Liu, Q. Zhang, L. Wang, J. Xiao, Q. Zhang, G. Luo, Z. Li, 
J. He, Y. Zhang, W. Ma, Plos Neg. Trop. Dis. {\bf 11} e0005973 (2017). 

\bibitem{Panja2023}
M. Panja, T. Chakraborty, S.S. Nadim, I. Ghosh, U. Kumar, N. Liu, 
Chaos, Sol. and Frac. {\bf 167} 113124 (2023).

\bibitem{Zhao2023}
X. Zhao, K. Li, C.K.E. Ang, K.H. Cheong, Chaos, Sol. and Frac. {\bf 168} 
113170 (2023).

\bibitem{Cabrera2022}
M. Cabrera, J. Leake, J.  Naranjo-Torres, N. Valero, J.C. Cabrera, 
A.J. Rodr\'iguez-Morales, Trop. Med. and Inf. Dis. {\bf 7} 322 (2022).

\bibitem{Francisco2021} 
M.E. Francisco, T.M. Carvajal, M. Ryo, K. Nukazawa, D.M. Amalin, 
K. Watanabe, Sci. of the total Env. {\bf 792} 148406 (2021).

\bibitem{Salim2021} 
N.A.M. Salim, Y.B. Wah, C. Reeves, M. Smith, W.F.W. Yaacob, R.N. Mudin, 
R. Dapari, N.N.F.F. Sapri, U. Haque, Sci. Rep. {\bf 11} 939 (2021).

\bibitem{Rahman2021} 
M.S. Rahman, C. Pientong, S. Zafar, T. Ekalak-Sananan, R.E. Paul, 
U. Haque, J. Rocklov, H.J. Overgaard, One Health {\bf 13} 100358 (2021).

\bibitem{Ochida2022} 
N. Ochida, M. Mangeas, M. Dupont-Rouzeyrol, C. Dutheil, C. Forfait, 
A. Peltier, E. Descloux, C. Menkes, Env. Health {\bf 21} 20 (2022).

\bibitem{Ming2022}
D.K. Ming, N.M. Tuan, B. Hernandez, S. Sangkaew, N.L. Vuong, H.Q. Chanh, 
N.V.V. Chau, C.P. Simmons, B. Wills, P. Georgiou, A.H. Holmes, S. Yacoub, 
Fron. in Dig. Health {\bf 4} (2022).


\bibitem{Roster2022}
K. Roster, C. Connaughton, F.A. Rodrigues, Amer. J. of Epid. {\bf 191} 1803-1812 (2022).

\bibitem{Ong2018}
J. Ong, X. Liu, J. Rajarethinam, S.Y. Kok, S. Liang, C.S. Tang, A.R. Cook, 
L.C. Ng, G. Yap, Plos Neg. Trop. Dis. {\bf 12} e0006587 (2018).

\bibitem{Benedum2020}
C.M. Benedum, K.M. Shea, H.E. Jenkins, L.Y.
Kim, N. Markuzon, Plos Neg. Trop. Dis. {\bf 14} e0008710 (2020).

\bibitem{Breiman2001}
L. Breiman, Mach. Lear. {\bf 45} 5-32 (2001).

\bibitem{Gendriz2022}
I. Sanchez-Gendriz, G.F.  de Souza, I.G.M. de Andrade, 
A.D.D. Neto, A.M. Tavares, D.M.S. Barros, A.H.F. de Morais, 
L.J. Galv\~ao-Lima, R.A.M. Valentim, {Sci. Rep.} {\bf 12} 6550 (2022).

\bibitem{Metereologia}
Instituto Nacional de Meteorologia: Minist\'erio da Agricultura e Pecu\'aria.
Available on: https://portal.inmet.gov.br Accessed on:  10 August 2023.

\bibitem{IquitosDados}
National Oceanic and Atmospheric Administration. 
Dengue Forecasting Project Data Repository. 
Available on: https://dengueforecast ing.noaa.gov 
Accessed on:  10 August 2023.

\bibitem{Sivigila}
Sistema Nacional de Vigilancia en Salud Pública -- Sivigila. 
Available on: https://portalsivigila.ins.gov.co 
Accessed on:  10 August 2023.

\bibitem{Trujillo2018}
J.C. Trujillo, H. Peter H, {Mend. Dat} {\bf V1} (2018).

\bibitem{SidneyGithub}
ClimateDengueForecast. Available on: https://github.com/ecgabrick/ClimateDengueForecast 
Accessed on: 22 March 2024. 

\bibitem{urca}
urca: Unit Root and Cointegration Tests for Time Series Data. 
Available on: https://CRAN.R-project.org/package=urca 
Accessed on: 21 March 2024.

\bibitem{Biau2016}
G. Biau, E. Scornet, {TEST} {\bf 25} 197--227 (2016). 

\bibitem{scikit-learn}
F. Pedregosa, G. Varoquaux, A. Gramfort, V. Michel, B. Thirion, 
O. Grisel, M. Blondel, P. Prettenhofer, R. Weiss, V. Dubourg, 
J. Vanderplas, A. Passos, D. Cournapeau, M. Brucher, M. Perrot, 
E. Duchesnay, {J. of Mach. Lear. Res.} {\bf 12} 2825--2830 (2011).

\bibitem{pdsr_r}
Use Time Series to Generate and Compare Power Spectral Density (PDSR). 
Available on: https://CRAN.R-project.org/package=psdr 
Accessed on: 1 March 2024. 

\bibitem{Forests}
A. Cutler, D.R. Cutler, J.R. Stevens JR,  Random Forests. In: Zhang c, MaY. (EDS)
Ensemble Machine Learning. (Springer, New York, 2012).

\bibitem{particao}
A.J. Izenman, Modern Multivariate Statistical Techniques. (Sprin\-ger, New York, 2008).
\end{thebibliography}
\end{document}